**Title: Lung Cancer in Argentina: A Modelling Study of Disease and Economic Burden**

**Authors**

**Andrea Alcaraz, MD MSc**

Department of Health Technology Assessment and Health Economics, Institute for Clinical Effectiveness and Health Policy (IECS), Buenos Aires, Argentina

Email: aalcaraz@iecs.org.ar

ORCID: 0000-0002-4260-8239

**Federico Rodriguez Cairoli, MD.**

Department of Health Technology Assessment and Health Economics, Institute for Clinical Effectiveness and Health Policy (IECS), Buenos Aires, Argentina

Email: fcairoli@iecs.org.ar

ORCID: 0000-0002-3029-4439

**Carla Colaci, MD.**

Department of Health Technology Assessment and Health Economics, Institute for Clinical Effectiveness and Health Policy (IECS), Buenos Aires, Argentina

Email: ccolaci@iecs.org.ar

ORCID: 0000-0002-5397-7531

**Constanza Silvestrini, BSc.**

Department of Health Technology Assessment and Health Economics, Institute for Clinical Effectiveness and Health Policy (IECS), Buenos Aires, Argentina

Email: csilvestrini@iecs.org.ar

ORCID: 0000-0001-8954-0991

**Carolina Gabay, MD**

Independent consultant in oncology, Buenos Aires, Argentina

Email: caro_gabay@hotmail.com

ORCID: 0009-0003-4966-4190

**Natalia Espinola, BSc., MSc.**

Department of Health Technology Assessment and Health Economics, Institute for Clinical Effectiveness and Health Policy (IECS), Buenos Aires, Argentina

Email: nespinola@iecs.org.ar

ORCID: 0000-0001-5511-3561

**Funding**

This study was made possible by the support of AstraZeneca Argentina S.A. The sponsor of the study had no role in the study design, data collection, data analysis, data interpretation or writing the manuscript.

**Competing interest**

The authors have declared that no competing interests exist.

**Availability of data and material**

All data involved in this study are included in the main manuscript and its supplementary information files.

**Ethical approval**

Not applicable

**Consent for publication**

Not applicable

**Authors contribution**

Concept and design, *AA, NE, FC*. Formal analysis, *FC, CC, CS*. Lung cancer expert, *CG*. General coordinator, *AA*. Review, interpretation and discussion of the results, *AA, NE, FC, CC, CS, CG*. Writing, reviewing and editing the manuscrip, *AA, NE, FC, CC, CS, CG*.

**Acknowledgement**

None.

**Corresponding author**

**Andrea Alcaraz, MD MSc**

Department of Health Technology Assessment and Health Economics, Institute for Clinical Effectiveness and Health Policy (IECS), Buenos Aires, Argentina

Doctor Emilio Ravignani 2024

Buenos Aires - Argentina

Email: [aalcaraz@iecs.org.ar](mailto:aalcaraz@iecs.org.ar)

Phone number: +54 9 221 523 0188

**Abstract**

*Objectives:* Lung cancer remains a significant global public health challenge and is still one of the leading cause of cancer-related death in Argentina. This study aims to assess the disease and economic burden of lung cancer in the country.

*Study design:* Burden of disease study

*Methods.* A mathematical model was developed to estimate the disease burden and direct medical cost attributable to lung cancer. Epidemiological parameters were obtained from local statistics, the Global Cancer Observatory, the Global Burden of Disease databases, and a literature review. Direct medical costs were estimated through micro-costing. Costs were expressed in US dollars (US$), April 2023 (1 US$ =216.38 argentine pesos). A second-order Monte Carlo simulation was performed to estimate the uncertainty.

*Results:* Considering approximately 10,000 deaths, 12,000 incident cases, and 14,000 5-year prevalent cases, the economic burden of lung cancer in Argentina in 2023 was estimated to be US$ 556.20 million (396.96 -718.20), approximately 1.4% of the total healthcare expenditure for the country. The cost increased with a higher stage of the disease and the main driver was the drug acquisition (80%). 179,046 Disability-adjusted life years could be attributable to lung cancer representing the 10% of the total cancer.

*Conclusion:* The disease and economic burden of lung cancer in Argentina implies a high cost for the health system and would represent 19% of the previously estimated economic burden for 29 cancers in Argentina.

**Background**

Lung cancer (LC) is still one of the most important causes of death in the world, in 2020 the World Health Organization (WHO) ranked it as the second leading cause of death. In terms of new cases, the WHO estimates about 2.21 million and 1.80 million of death by 2020 worldwide.[1] The Global Burden of Disease (GBD) study, in 2019 estimate that there were 45.9 million (95% CI 42.3 - 49.3) DALYs due to tracheal, bronchus, and LC, of which 98.8% came from Years of Life Lost (YLLs) and 1.2% from Years Lived with Disability (YLDs).[2]

Piñeros et al. reported the updated profile of cancer burden, patterns and trends in Latin America and the Caribbean.[3] The report showed that the highest incidence and mortality rates were observed in Uruguay, Cuba and Argentina in both men and women. LC ranks as the fourth most commonly diagnosed cancer in South America and the Caribbean, leading cause of cancer deaths in both sexes (12%).[3]

In Argentina, LC registered the highest number of deaths, accounting for almost 15% of all cancer deaths by 2020.[4] In women, LC ranked fifth, following breast cancer, cervical cancer, colorectal cancer, and uterine cancer (body), representing 3.2% of all cancer cases; while in men, it ranked second, right after colorectal cancer, representing 10.3% of all cancer cases, in the period 2012-2020, according to The Institutional Registry of Tumours in Argentina (RITA) report.[5]

Tobacco remains the leading risk factor for the development of LC. In the year 2023, Pichon -Riviere A, Bardach A, Cairoli, et al published the results of the disease burden attributable to tobacco in eight Latin American countries.[6] Argentina had the highest relative burden of disease and economic burden attributable to tobacco use, standing out with the highest percentage of LC cases attributable to tobacco at 19.2%, surpassing Chile and Brazil with rates of 16.8% and 15.1%, respectively. Tobacco was responsible for 19.2% of deaths due to LC in the country.

LC can be broadly categorised into two main types: non-small cell LC (NSCLC), and small cell LC (SCLC).[7] Based on its extension NSCLC can be classified in 4 stages, with stage IV being the most aggressive. Stage I and II current treatments are mainly based on surgical resection of the tumour and on the use of radiotherapy. For stage II and III systematic treatments (chemotherapy, immunotherapy) have been shown to improve 5-year overall survival.[8] Most of the patients with NSCLC have advanced disease at diagnosis (stage IV), in which the determination of treatment options depends on biopsy and the presence of certain biomarkers to determine mutations and guide treatment.[9] SCLC is characterised by its rapid growth, tendency to metastazise and low survival rates. SCLC is classified into two stages: limited stage, in which the disease is limited to the hemithorax and the treatment consists of chemoradiotherapy (QT) followed by prophylactic cranial irradiation. On the other hand, extensive stage SCLC is usually treated with chemoimmunotherapy, with or without consolidation radiation.[10] While, SCLC is characterised by a rapid response to chemotherapy (QT) and sensitivity to radiotherapy (RT), due to resistance to treatment, its overall survival (OS) at 5 years is considered low. SCLC is equally prevalent in men and women.[10]

LC is a prevalent disease with a high incidence rate, often diagnosed at an advanced stage, resulting in an approximate mortality rate of 50%. The introduction of new treatments and personalized medicine has led to an increase in the survival rates, but it has also resulted in rising costs. This is the first study in Argentina that comprehensively assesses both aspects: the disease and economic burden of LC. The aim of this study was to estimate the disease and economic burden of LC in Argentina for the year 2023.

The analysis was assessed by cancer type, stage, sex, and healthcare sectors in the country (public, social security, and private sectors).

**Methods**

*Model structure*

A mathematical model was developed to estimate the disease and economic burden attributable to LC in Argentina (See Figure 1). The model estimates the total number of cases and deaths, and the total direct medical costs disaggregated by type of LC, stage and sex for the year 2023. Total LC patients are estimated and taking into account all incidents cases (new LC diagnosed in a year) and all prevalent cases. For prevalent cases, all those patients with a previous diagnosis of LC between the years 2018 and 2022, who are still alive in the year 2023 were considered. Additionally, the number of deaths among incident and prevalent patients were estimated. The total direct medical costs were estimated by incident and prevalent, type of LC, stage and the three healthcare sectors. In addition, the model estimated the DALYs attributable to the disease, which were obtained from the sum of the YLLs and the YLDs.

**Figure 1.** Model framework

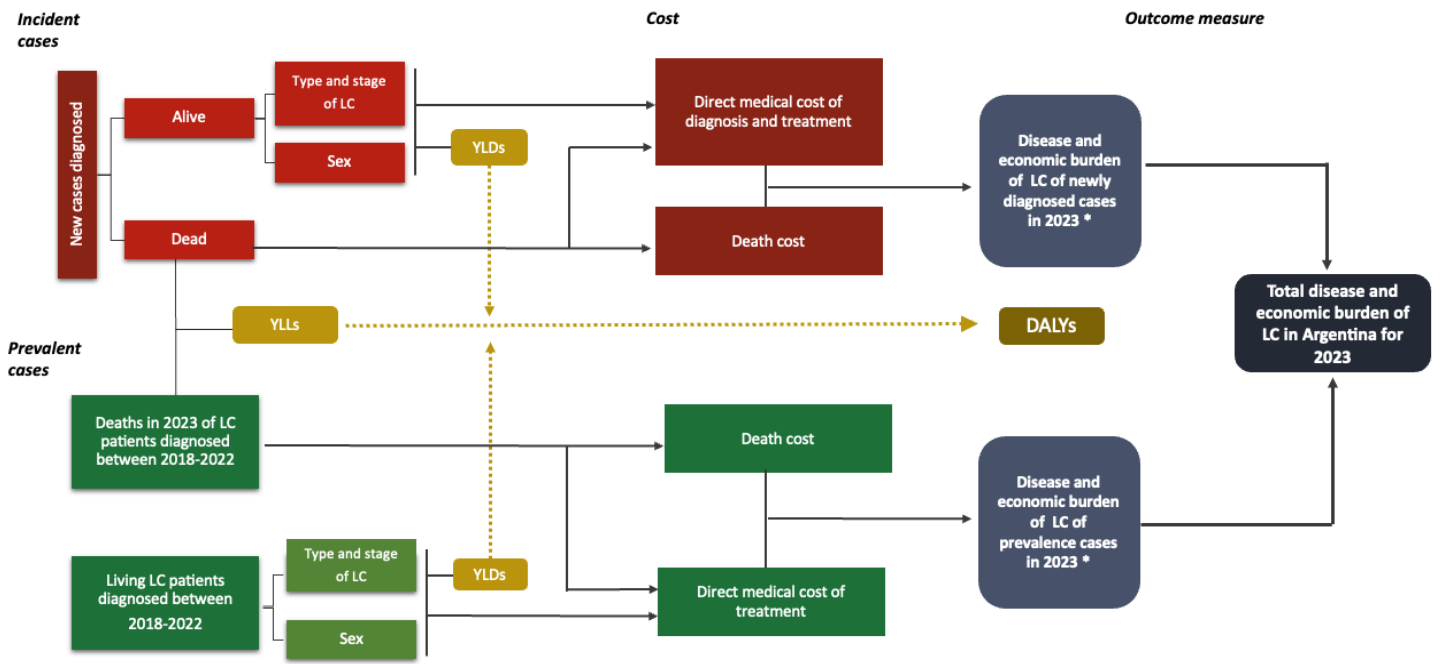

*Abbreviations:* **LC:** lung cancer, **YLDs:** years of health life lost due to disability, **YLLs:** years of life lost, **DALYs:** disability-adjusted life years

*Epidemiological data*

The data used to populate the model was obtained from global observatories, literature review and validation by a local expert in LC. The total Argentinian population was obtained from the latest national census in 2022.[11] Information on the number of incident cases, prevalence cases (1, 3, and 5-year prevalence), and deaths due to LC per sex in 2020 were collected by Global Cancer Observatory (Globocan) for Argentina.[12] **Table 1** shows the details on these parameters.

The distribution of incident cases by cancer type and stage were obtained from Thai et al. (2021),[7] and data published by the Australian Institute of Health and Welfare,[13] respectively, given the lack of local sources with accurate information. All these values were validated by the local expert (**Table 1**)

To obtain the distribution by cancer type and stage of prevalent patients, we modelled the survival of incident cases until five years of diagnosis including information on specific survival probabilities by cancer type, stage, and year since diagnosis. Prevalence cases beyond five-years were not included by considered marginal (expert consensus).

Once the distribution of incident cases was determined, we employed the one-year survival curves from Chansky et al. (2017)[14] to model the distribution of survivors based on LC type, stage, and sex, adjusted to fit the one-year overall survival probability and absolute number of cases from Globocan[12] (**Supplementary Material [SM] Table S1**). The same procedure was done, using the two, three, four, and five-year survival probabilities reported by Chansky et al[14], until obtaining the distribution of the 5-year prevalent cases (**SM Table S2**).

The number of deaths attributed to LC in Argentina was collected from Globocan.[12] adjusted to differentiate between incident and prevalent cases.

DALYs were calculated from YLLs and YLDs. YLLs were estimated using Argentinian life expectancy tables by sex from WHO online Global Health Observatory data repository.[15] YLDs were estimated for the incident and prevalence cases using disease-stage specific disability weights (DWs) from the GBD study.[16] For localized stages of incident cases, we considered the DW value of 0.288 (diagnosis and primary therapy phase). For localized stages of LC prevalent cases, we considered the DW of 0.049 as a conservative approach (controlled phase). Finally, we used the DW of 0.451 (metastatic phase) for stage IV/extended stage.

**Table 1.** Main epidemiological input parameters.

| General population data | Data | Source |
|---|---|---|
| Total population of Argentina (2022) | 46.234.830 | Argentine National Census, 2022[11] |
| Total female population (2022) | 23.525.352 | Argentine National Census, 2022[11] |
| Total male population (2022) | 22.709.478 | Argentine National Census, 2022[11] |
| Total new cases of lung cancer (incident) | 12.110 | Globocan Argentina, 2020 [17] |
| Total number of LC patients (1-year prevalent cases) | 6.443 | Globocan Argentina, 2020 [17] |
| Total number of LC patients (5-year prevalent cases) | 14.103 | Globocan Argentina, 2020 [17] |
| Total deaths in one year due to LC | 10.729 | Globocan Argentina, 2020 [17] |
| Mortality:Incidence Ratio | 0.88 | Rajesh Sharma et al, 2022 [18] |
| **Epidemiology data on LC at diagnosis** | | |
| *Non-Small Cell Lung Cancer* | **85%** | Thai et al, 2021[7] |
| Stage I | 12,5% | Australian Institute of Health and Welfare, 2019 [13] |
| Stage II | 7,5% | Australian Institute of Health and Welfare, 2019 [13] |
| Stage III | 25% | Australian Institute of Health and Welfare, 2019 [13] |
| Stage IV | 55% | Australian Institute of Health and Welfare, 2019 [13] |
| *Small Cell Lung Cancer* | **15%** | Thai et al, 2021 [7] |
| Limited (Stage I, II, III) | 35% | Australian Institute of Health and Welfare, 2019 [13] |
| Extended (Stage IV) | 65% | Australian Institute of Health and Welfare, 2019 [13] |
| **Epidemiology data on LC 5 years prevalent patients** | | |
| *Non-Small Cell Lung Cancer* | **87%** | Estimated |
| Stage I | 27.73% | Estimated |
| Stage II | 12.19% | Estimated |
| Stage III | 28.19% | Estimated |
| Stage IV | 31.89% | Estimated |
| *Small Cell Lung Cancer* | **13%** | Estimated |
| Limited (Stage I, II, III) | 68.82 | Estimated |
| Extended (Stage IV) | 31.18 | Estimated |

*Abbreviation: LC: lung cancer*

*Economic data*

We used a micro-costing approach to estimate the annual direct medical costs. The identification, rate of use and measurement of health resources used for the disease management (phases of diagnosis, treatment and follow-up) were estimated by literature review and local expert opinion. The cost categories included in the model were diagnosis, surgery, radiotherapy and drugs (acquisition and administration), follow-up and management of adverse events. The main healthcare resources considered were medical consultations, laboratory tests, imaging tests, hospitalisation and drugs. The healthcare system in Argentina is fragmented into three sectors: public, social security and private.[19] The social security is the largest and provides healthcare coverage to 46% of the Argentine population, 16% are covered by the private sector, and 38% by the public sector.[20] This translates into different healthcare resource costs by sector, then direct medical costs were estimated for each of the sectors. To obtain the direct medical costs for the health system, an average weighted by coverage rates of each sector was used. All costs were estimated in Argentinian pesos (ARS) and then converted to US dollars (US$) for April 2023 (1 US$ = 216.38 ARS).[21]

The unit cost of healthcare resources by health sector were obtained from IECS unit cost database.[22] **SM Tables S6, S7, S8.**

The drug acquisition costs were obtained from public databases that provide the retail prices of drugs in Argentina[23] and converted to wholesale prices using the conversion factor recommended by the Argentinian Ministry of Economy (referred to as the ex-factory price).[24] We assumed that the drug acquisition costs are the same for the three perspectives of analysis. The therapies included for the treatment of LC, both small cell and non-small cell, were anaplastic lymphoma kinase inhibitors (ALK), tyrosine kinase inhibitors (antiEGFR) drugs, immunotherapy, and chemotherapy. **SM Tables S9, S10, S11, S12**.

We included those adverse events (AEs) with a score greater than 3 on the Common Terminology Criteria for Adverse Events V 5.0 (CTCAE)[25] scale, and a prevalence rate greater than 3% . We considered: anaemia, neutropenia, thrombocytopenia, pneumonia, diarrhoea, vomiting, nausea, and pneumonitis.[26] [27–35] **SM Tables S13, S14**.

The cost of death estimated for incidents is composed of days of hospitalization, and the estimated cost of death for prevalents is composed of half a year of treatment in their respective stages, palliative care, and days of hospitalization. The cost of palliative care was obtained from Lamfre et al. (2023).[36] **SM Tables S15, S16**.

*Uncertainty analysis*

A second-order Monte Carlo simulation was conducted using Excel version 16.75. The variables considered were: absolute number of deaths, costs per patient per year, and the mortality:incidence (m:i) ratio. Normal distributions were assumed for the three parameters. For the number of deaths, we considered a variation of +/- 13% relative to the central value, based on national death data from the 2019 National Statistics (DEIS)[37], and previous GBD study reports (2019).[38] For the cost per patient per year, we considered a variation of +/- 25%, following usual variation for the country.[39] Finally, for the uncertainty of the mortality:incidence ratio, we calculated a possible range of the m:i ratio of 0.82-0.94 (0.88 for South American countries), based on Rajesh Sharma et al.[18]
.

**RESULTS**

It was estimated 12,110 new cases, 14,103 5-year prevalent cases, and 10,729 deaths due to LC in a year in Argentina.[12] Table 1 presented the distribution percentages of different types of cancer and stages for incident and prevalents patients. Additionally, 13% of prevalent cases were diagnosed with SCLC, with 68.8% in the limited stage and 31.2% in the extended stage.

Regarding deaths, estimates showed that 53% of these deaths (5.667) were among incidents, and the remaining 5,062 deaths in prevalent patients.

Finally, the estimated YLLs were found to be 69,040 years for women and a higher value of 102,908 years for men. Furthermore, the estimated YLDs resulted in 2,629 years for women and 4,468 years for men. These translate into 179,046 DALYs. (**SM Table S3, S4, S5**)

The average annual cost per patient with LC was higher in incident patients (US$22,889) than in prevalent patients (US$15,990), and in both cases, it increased as the disease progressed from early to advanced stages (Figure 2). **SM Tables S17, S18.**

**Figure 2. Average annual cost per patient due to lung cancer by cancer type and stage for healthcare system in**

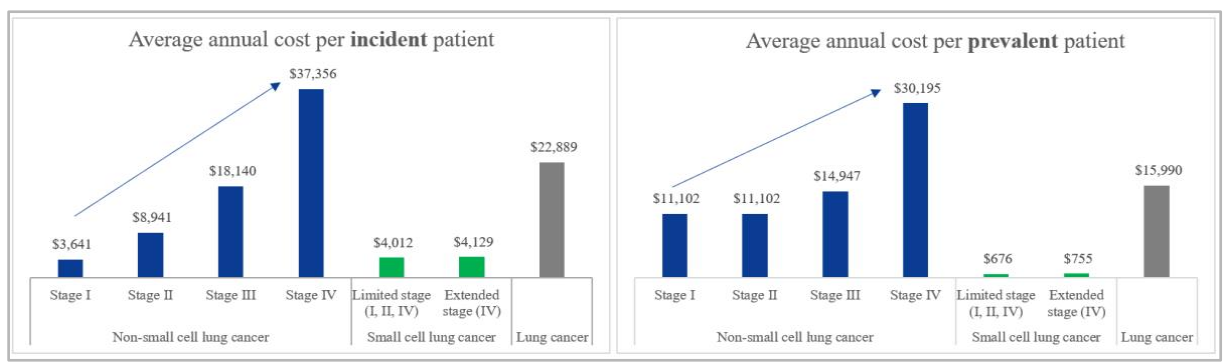

**Argentina (April 2023)**
*Note: Costs are reported in 2023 US$ dollars.*

Figure 3 shows the composition of the average annual cost. The main component of annual cost was the drugs costs in the majority of stages, except for incident stage I of NSCLC and prevalent limited stage in SCLC, where the surgery and expenses related to medical consultations, procedures, treatments, and laboratory tests are the main components, respectively.

**Figure 3. Composition of the average annual cost per lung cancer patient in Argentina (April 2023)**

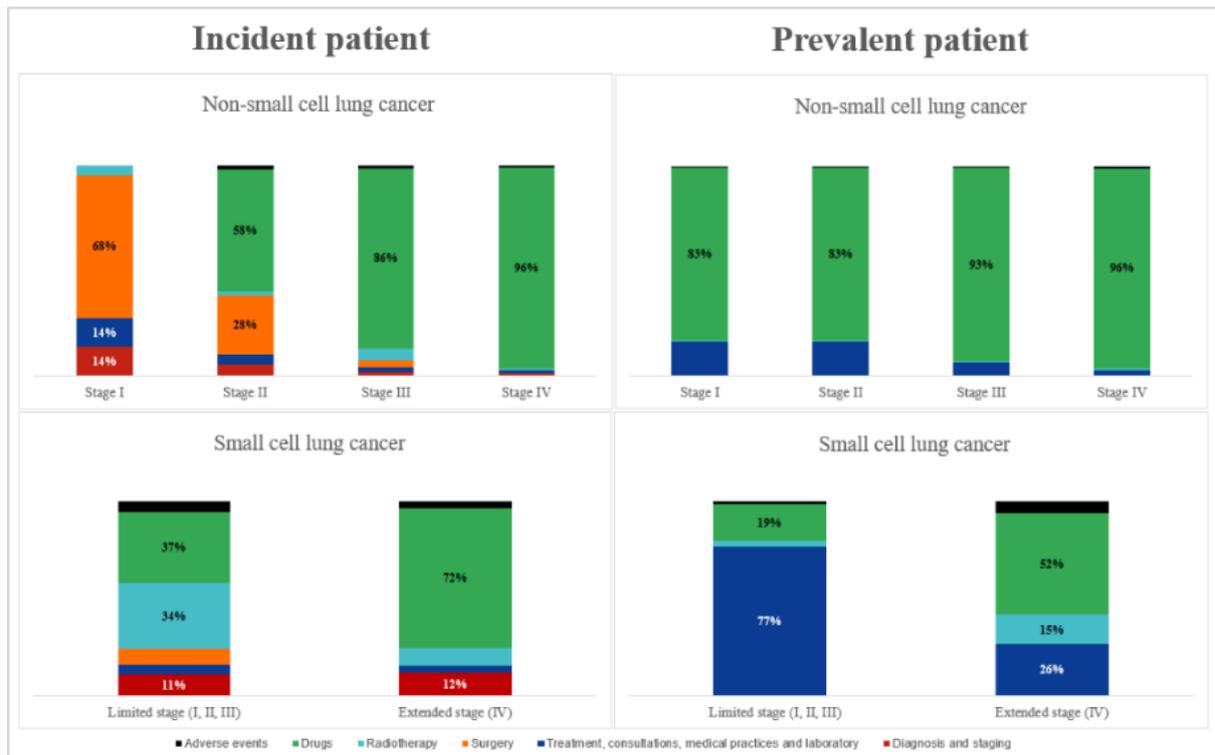

Finally, the cost of death was estimated in US$969.43 per average incident death for both cancer types, while the estimated cost of death for prevalent cases with NSCLC was US$5,778.10, US$7,316.14, and US$13,415.35 for stages I-II, III, and IV, respectively. In addition, the estimated cost of death for prevalent cancer with SCLC was calculated at US$1,607.87 and US$1,639.29 in limited and extended stages, respectively. **SM Tables S19, S20.**

*Total Annual disease and direct medical cost attributable to LC in Argentina*

Table 2 shows the results of the average total cost of LC per type of cancer, stage, and sex. The average annual cost of LC was estimated in 556.20 million, where US$282.68 million are due incident cases (50.8%), US$226.75 million (40.8%) are due prevalent cases, and US$46,76 million (8.4%) are due death costs. The average annual cost of LC was US$203.78 million for the public sector, US$260.10 million for the social security sector, and US$92.31 million for the private sector (**SM Table S21**). In addition, the results showed that the total cost of LC was higher in men (US$180.63 million for NSCLC, and US$140.51 million for SCLC) compared to women (US$102.05 million for NSCLC and US$86.24 million for SCLC).

**Table 2.** Disease and economic burden attributable to lung cancer in Argentina for 2023. Costs are expressed in US$ for 2023.

| | Cases/Death | | | Total weighted cost million US$ | | |
|---|---|---|---|---|---|---|
| | Men | Women | Total | Men | Women | Total |
| | n | n | n | | | |
| **Incidents** | | | | | | |
| **Non Small Cells Lung Cancer** | | | | | | |
| Stage I | 822 (703 - 945) | 464 (397 - 534) | 1,287 (1,100 - 1,479) | $2.99 ($2.13 - $3.87) | $1.70 ($1.21- $2.18) | $4.68 ($3.34 - $6.05) |
| Stage II | 493 (422 - 567) | 279 (238 - 320) | 772 (660 - 888) | $4.41 ($3.15 - $5.70) | $2.50 ($1.78 - $3.22) | $6.90 ($4.93 - $8.91) |
| Stage III | 1,644 (1,405 - 1,890) | 929 (794 - 1,068) | 2,573 (2,199 - 2,958) | $29.83 ($21.29 - $38.52) | $16.85 ($12.03 - $21.76) | $46.68 ($33.32 - $60.28) |
| Stage IV | 3,618 (3,091 - 4,159) | 2,044 (1,745 - 2,350) | 5,661 (4,838 - 6,508) | $135.14 ($96.45 - $174.50) | $76.35 ($54.49 - $98.59) | $211.49 ($150.94 - $273.09) |
| **Small Cells Lung Cancer** | | | | | | |
| Limited (Stage I, II, III) | 406 (347 - 467) | 230 (196 - 264) | 636 (543 - 731) | $1.63 ($1.16 - $2.10) | $0.92 ($0.66 - $1.19) | $2.55 ($1.82 - $3.29) |
| Extended (Stage IV) | 754 (645 - 867) | 426 (364 - 490) | 1,181 (1,010-1,357) | $3.11 ($2.22 - $4.02) | $1.76 ($1.26 - $2.27) | $4.87 ($3.48 - $6.29) |
| **Death** | 3,621 (3,150 - 4,094) | 2,046 (1,779 - 2,313) | 5,667 (4,929 - 6,407) | $3.51 ($2.50- $4.51) | $1.99 ($1.41 - $2.55) | $5.49 ($3.92 - $7.06) |
| **Subtotal** | 7,738 (6,613 - 8,896) | 4,372 (3,736 - 5,026) | 12,110 (10,349 - 13,922) | $180.63 ($128.92- $233.24) | $102.05 ($72.83 - $131.77) | $282.68 ($201.75 - $365.016) |
| **Prevalents** | | | | | | |
| **Non Small Cells Lung Cancer** | | | | | | |
| Stage I | 2,098 (1,793 - 2,412) | 1,323 (1,130 - 1,521) | 3,421 (2,923 - 3,933) | $23.29 ($16.63 - $30.07) | $14.69 ($10.48 - $18.96) | $37.98 ($27.11 - $49.03) |
| Stage II | 927 (792 - 1,065) | 577 (493 - 663) | 1,504 (1,285 - 1,729) | $10.29 ($7.34 - $13.28) | $6.41 ($4.57 - $8.27) | $16.69 ($11.92 - $21.55) |
| Stage III | 2,149 (1,836 - 2,470) | 1,329 (1,135 - 1,527) | 3,477 (2,971 - 3,997) | $32.12 ($22.92 - $41.46) | $19.85 ($14.17 - $25.62) | $51.96 ($37.09 - $67.08) |
| Stage IV | 2,452 (2,096 - 2,819) | 1,484 (1,268 - 1,707) | 3,937 (3,364 - 4,526) | $74.05 ($52.86 - $95.60) | $44.82 ($31.99 - $57.86) | $118.88 ($84.86 - $153.46) |
| **Small Cells Lung Cancer** | | | | | | |
| Limited (Stage I, II, III) | 747 (638 - 859) | 467 (399 - 537) | 1,214 (1,037 - 1,395) | $0.51 ($0.36 - $0.65) | $0.32 ($0.23 - $0.41) | $0.82 ($0.59 - $1.06) |
| Extended (Stage IV) | 345 (295 - 396) | 207 (177 - 238) | 552 (472 - 634) | $0.26 ($0.19 - $0.34) | $0.16 ($0.11 - $0.20) | $0.42 ($0.30 - $0.54) |
| **Subtotal** | 8,718 (7,450 - 10,022) | 5,386 (4,603 - 6,192) | 14,103 (12,052 - 16,213) | $140.51 ($100.30 - $181.39) | $86.24 ($61.56 - $111.32) | $226.75 ($161.86 - $292.71) |
| **Death after one year of diagnosis** | 3,415 (2,970 - 3,861) | 1,647 (1,433 - 1,862) | 5,062 (4,403 - 5,724) | $31.55 ($22.49 - $40.54) | $15.22 ($10.85 - $19.55) | $46.76 ($33.34 - $60.09) |
| **Total cost of lung cancer in Argentina** | - | - | - | $352.69 ($251.72 - $455.43) | $203.50 ($145.24 - $262.77) | $556.20 ($396.96 - $718.20) |

**DISCUSSION**

LC remains a leading cause of death worldwide. In Argentina, it ranked third in 2020, with approximately 12,000 new cases diagnosed each year, and more that 10,000 deaths, of which 57% are attributable to patients diagnosed within that same year. That represented a loss of 179,046 DALYs. The total average cost of LC amounts to $556 million dollars in direct medical costs, representing approximately 1.4% of the entire healthcare expenditure in Argentina.

In the study, the economic burden of LC is 0.09% of GDP. According to a recently published study conducted in 204 countries for 29 cancers, the macroeconomic cost of LC is estimated at 0.08%[42] of GDP. In addition, the study estimated a macroeconomic burden of cancer as a percentage of GDP of 0.35% for Latin America and 0.46% for Argentina.[42]. In this sense, the economic burden of LC estimated in our study would represent 19% share of this estimated economic burden in Argentina. Additionally, a study by Pichon-Riviere in 2020 estimated the economic burden of LC for main countries in Latin America ranging from 0.03% in Honduras to 0.33% in Uruguay[43], while one study reported an economic burden of LC for the US of 0.04% of GDP.[44]

Our study reported a lower DALY burden from LC compared to GBD2 for Argentina 2019 (318,837.68 DALYs). This likely occurred because GBD estimated a higher number of annual deaths (13,879) compared to Globocan (10,729), making our results conservative. A study in Australia, the Philippines and Singapore, in line with our results, estimated that DALYs made up the majority of LC DALYs (95%).[40] This highlights the significant impact of LC on global mortality, as according to GBD LC is the third leading cause of total DALYs (2.54%), behind ischaemic heart disease (6.21%) and stroke (4.24%).[41]

Despite the methodological heterogeneity inherent in LC cost studies, which impedes direct comparison of results across studies, a literature review facilitated the observation of a certain similarity in the relationship between NSCLC and SCLC costs[45,46,47,48,49] and in the relationship of stages within each type of cancer.[47,48] Significant variability was observed in annual direct medical costs per patient, expressed as a percentage of GDP per capita. NSCLC costs ranged in a lower range than estimated in our study (27%[48]-106%[46] vs. 163%), whereas SCLC costs ranged in a higher range than estimated in our study (27%[48]-95%[46] vs. 17%).

In the study, acquisition costs constitute a significant part of the overall cost (76% for incident and 82% for prevalent) and increase from early to late stages. This finding diverges from similar studies conducted in high-income countries, where hospitalization primarily drives costs.[48,49] This difference could be attributed to methodological differences, variations in healthcare resources utilization rates, disparities in relative prices of healthcare resources and drugs, as well as disparities in drug accessibility across countries.

Our study has the following strengths: best available epidemiological sources and epidemiological and cost information by cancer type, stage, sex and perspective from the different sub-sectors of the Argentinian health system. However, it is important to point out the limitations of our study: costs estimated using the micro-costing method with the guidance of clinical experts due to the lack of local data; the same distribution of the LC population by health subsector was assumed as the coverage rate of each subsector in Argentina; indirect costs (lost labour productivity, unpaid informal care, among others) that would probably increase the economic impact were not included.[49] According to Cicin et al. (2021)[49] in Turkey, indirect costs accounted for 69% of the overall economic burden of LC.

Given the rising incidence of LC, countries need to be prepared keeping translating research into real-world practice, enhancing prevention policies, supporting equitable access to healthcare, improving the effectiveness of healthcare services, and increasing public awareness.[50] The WHO recognises the substantial impact of LC on global health and has undertaken initiatives centred around tobacco control, cancer prevention, early detection, and enhancing access to high-quality cost effective treatment.[51]

**Conclusion**

The disease and economic burden of LC in Argentina implies a high cost for the health system. More efficient prevention programs and control policies are needed.

**REFERENCES**

1. Cancer. World Health Organization. Published February 3, 2022. Accessed May 2024. https://www.who.int/news-room/fact-sheets/detail/cancer

2. Ebrahimi H, Aryan Z, Saeedi Moghaddam S, et al. Global, regional, and national burden of respiratory tract cancers and associated risk factors from 1990 to 2019: a systematic analysis for the Global Burden of Disease Study 2019. *The Lancet Respiratory Medicine*. 2021;9(9):1030-1049. doi:10.1016/S2213-2600(21)00164-8

3. Piñeros M, Laversanne M, Barrios E, et al. An updated profile of the cancer burden, patterns and trends in Latin America and the Caribbean. *Lancet Reg Health Am*. 2022;13:None. doi:10.1016/j.lana.2022.100294

4. Natalia Aráoz, Olivos Gisel Fattore, Cecilia Maturo. *Cancer Mortality Bulletin in Argentina*. Ministry of Health ; 2020.

5. Jose Enrique Carrizo Olalla, Gisel Fattore, Cecilia Maturo. The Institutional Registry of Tumors in Argentina 2012-2020. Published May 27, 2021. Accessed July 26, 2023. https://bancos.salud.gob.ar/sites/default/files/2023-02/2023-02-Bolet%C3%ADn-RITA-v-f.pdf

6. Pichon-Riviere A, Bardach A, Cairoli FR, et al. Health, economic and social burden of tobacco in Latin America and the expected gains of fully implementing taxes, plain packaging, advertising bans and smoke-free environments control measures: a modelling study. *Tob Control*. Published online 2023. https://tobaccocontrol.bmj.com/content/early/2023/05/24/tc-2022-057618.abstract

7. Thai AA, Solomon BJ, Sequist LV, Gainor JF, Heist RS. Lung cancer. *Lancet*. 2021;398(10299):535-554. doi:10.1016/S0140-6736(21)00312-3

8. Postmus PE, Kerr KM, Oudkerk M, et al. Early and locally advanced non-small-cell lung cancer (NSCLC): ESMO Clinical Practice Guidelines for diagnosis, treatment and follow-up. *Ann Oncol*. 2017;28(suppl_4):iv1-iv21. doi:10.1093/annonc/mdx222

9. Planchard D, Popat S, Kerr K, et al. Metastatic non-small cell lung cancer: ESMO Clinical Practice Guidelines for diagnosis, treatment and follow-up. *Ann Oncol*. 2018;29(Suppl 4):iv192-iv237. doi:10.1093/annonc/mdy275

10. Dingemans AMC, Früh M, Ardizzoni A, et al. Small-cell lung cancer: ESMO Clinical Practice Guidelines for diagnosis, treatment and follow-up☆. *Ann Oncol*. 2021;32(7):839-853. doi:10.1016/j.annonc.2021.03.207

11. Statistics. Census 2022. National Institute of Statistics and Census (INDEC). Published January 31, 2023. Accessed February 9, 2023. https://www.indec.gob.ar/indec/web/Nivel3-Tema-2-18

12. Global Cancer Observatory. Globocan. International Agency for Research on Cancer. World Health Organization. Accessed February 10, 2023. https://gco.iarc.fr/

13. Relative survival by stage at diagnosis (lung cancer). National Cancer Control Indicators. Australian Government. Published April 1, 2019. Accessed February 10, 2023. https://ncci.canceraustralia.gov.au/outcomes/relative-survival-rate/relative-survival-stage-diagnosis-lung-cancer

14. Chansky K, Detterbeck FC, Nicholson AG, et al. The IASLC Lung Cancer Staging Project:

External Validation of the Revision of the TNM Stage Groupings in the Eighth Edition of the TNM Classification of Lung Cancer. *J Thorac Oncol*. 2017;12(7):1109-1121. doi:10.1016/j.jtho.2017.04.011

15. Life table by country. Expectation of life at age x. World Health Organization (WHO). Published December 6, 2020. Accessed February 14, 2023. https://www.who.int/data/gho/data/indicators/indicator-details/GHO/gho-ghe-life-tables-by-country

16. Global Burden of Disease Study 2019 (GBD 2019) Disability Weights. Accessed July 20, 2023. https://ghdx.healthdata.org/record/ihme-data/gbd-2019-disability-weights

17. World Health Organization. *International Agency for Research on Cancer. Globocan Argentina 2020.*; 2021. https://gco.iarc.fr/today/data/factsheets/populations/32-argentina-fact-sheets.pdf

18. Sharma R. Mapping of global, regional and national incidence, mortality and mortality-to-incidence ratio of lung cancer in 2020 and 2050. *Int J Clin Oncol*. 2022;27(4):665-675. doi:10.1007/s10147-021-02108-2

19. Palacios A, Espinola N, Rojas-Roque C. Need and inequality in the use of health care services in a fragmented and decentralized health system: evidence for Argentina. *Int J Equity Health*. 2020;19(1):67. doi:10.1186/s12939-020-01168-6

20. de la Nación M de S. Análisis de situación de salud: Republica Argentina. Published online 2018.

21. Central Bank of Argentina. Exchange rate by date. Central Bank of Argentina. Accessed December 2022. https://www.bcra.gob.ar/MediosPago/Tipos_de_Cambio_SML.asp

22. Palacios A, Balan D, Garay OU, et al. HT3 BASE DE COSTOS UNITARIOS EN SALUD EN ARGENTINA: UNA FUENTE DE INFORMACIÓN CONTINUAMENTE ACTUALIZADA PARA EVALUACIONES ECONOMICAS Y ANALISIS DE IMPACTO PRESUPUESTARIO EN UN SISTEMA DE SALUD FRAGMENTADO. *Value in Health Regional Issues*. 2019;19:S8. doi:10.1016/j.vhri.2019.08.042

23. Grupo AlfaBeta. Precio de medicamentos. Alfa Beta. Published 2020. Accessed April 2023. http://www.alfabeta.net/precio/

24. Garfinkel F, Méndez Y. *Value Chain Reports: Health, Pharmacy and Medical Equipment*. Office for Economic Policy and Development Planning; 2016.

25. Common Terminology Criteria for Adverse Events (CTCAE) Version 5.0. cancer.gov. Published March 15, 2023. Accessed November 27, 2017. https://ctep.cancer.gov/protocoldevelopment/electronic_applications/docs/ctcae_v5_quick_reference_5x7.pdf

26. Mok T, Camidge DR, Gadgeel SM, et al. Updated overall survival and final progression-free survival data for patients with treatment-naive advanced ALK-positive non-small-cell lung cancer in the ALEX study. *Ann Oncol*. 2020;31(8):1056-1064. doi:10.1016/j.annonc.2020.04.478

27. Ramalingam SS, Vansteenkiste J, Planchard D, et al. Overall Survival with Osimertinib in Untreated, EGFR-Mutated Advanced NSCLC. *N Engl J Med*. 2020;382(1):41-50. doi:10.1056/NEJMoa1913662

28. Sequist LV, Yang JCH, Yamamoto N, et al. Phase III study of afatinib or cisplatin plus

pemetrexed in patients with metastatic lung adenocarcinoma with EGFR mutations. *J Clin Oncol*. 2013;31(27):3327-3334. doi:10.1200/JCO.2012.44.2806

29. Antonia SJ, Villegas A, Daniel D, et al. Durvalumab after Chemoradiotherapy in Stage III Non-Small-Cell Lung Cancer. *N Engl J Med*. 2017;377(20):1919-1929. doi:10.1056/NEJMoa1709937

30. Herbst RS, Giaccone G, de Marinis F, et al. Atezolizumab for First-Line Treatment of PD-L1-Selected Patients with NSCLC. *N Engl J Med*. 2020;383(14):1328-1339. doi:10.1056/NEJMoa1917346

31. Hellmann MD, Paz-Ares L, Bernabe Caro R, et al. Nivolumab plus Ipilimumab in Advanced Non-Small-Cell Lung Cancer. *N Engl J Med*. 2019;381(21):2020-2031. doi:10.1056/NEJMoa1910231

32. Reck M, Rodríguez-Abreu D, Robinson AG, et al. Pembrolizumab versus Chemotherapy for PD-L1-Positive Non-Small-Cell Lung Cancer. *N Engl J Med*. 2016;375(19):1823-1833. doi:10.1056/NEJMoa1606774

33. Hanna N, Shepherd FA, Fossella FV, et al. Randomized phase III trial of pemetrexed versus docetaxel in patients with non-small-cell lung cancer previously treated with chemotherapy. *J Clin Oncol*. 2004;22(9):1589-1597. doi:10.1200/JCO.2004.08.163

34. Sandler A, Gray R, Perry MC, et al. Paclitaxel-carboplatin alone or with bevacizumab for non-small-cell lung cancer. *N Engl J Med*. 2006;355(24):2542-2550. doi:10.1056/NEJMoa061884

35. Borghaei H, Paz-Ares L, Horn L, et al. Nivolumab versus Docetaxel in Advanced Nonsquamous Non-Small-Cell Lung Cancer. *N Engl J Med*. 2015;373(17):1627-1639. doi:10.1056/NEJMoa1507643

36. Lamfre L, Hasdeu S, Coller M, Tripodoro V. Análisis de costo-efectividad de los cuidados paliativos a pacientes oncológicos de fin de vida. *Cad Saúde Pública*. 2023;39(2):ES081822. doi:10.1590/0102-311XES081822

37. Dirección de Estadísticas e Información de la Salud. Argentina.gob.ar. Published December 4, 2020. Accessed July 20, 2023. https://www.argentina.gob.ar/salud/deis

38. Vos T, Lim SS, Abbafati C, et al. Global burden of 369 diseases and injuries in 204 countries and territories, 1990–2019: a systematic analysis for the Global Burden of Disease Study 2019. *Lancet*. 2020;396(10258):1204-1222. doi:10.1016/S0140-6736(20)30925-9

39. Nuijten MJC, Mittendorf T, Persson U. Practical issues in handling data input and uncertainty in a budget impact analysis. *Eur J Health Econ*. 2011;12(3):231-241. doi:10.1007/s10198-010-0236-4

40. Morampudi S, Das N, Gowda A, Patil A. Estimation of lung cancer burden in Australia, the Philippines, and Singapore: an evaluation of disability adjusted life years. *Cancer Biol Med*. 2017;14(1):74-82. doi:10.20892/j.issn.2095-3941.2016.0030

41. Global Burden of Disease (GBD) Compare tables. Argentina. Institute for Health Metrics and Evaluation (IHME). Published 2019. Accessed September 12, 2023. https://vizhub.healthdata.org/gbd-compare/

42. Chen S, Cao Z, Prettner K, et al. Estimates and Projections of the Global Economic Cost of 29 Cancers in 204 Countries and Territories From 2020 to 2050. *JAMA Oncol*. 2023;9(4):465-472. doi:10.1001/jamaoncol.2022.7826

43. Pichon-Riviere A, Alcaraz A, Palacios A, et al. The health and economic burden of smoking in

12 Latin American countries and the potential effect of increasing tobacco taxes: an economic modelling study. *Lancet Glob Health*. 2020;8(10):e1282-e1294. doi:10.1016/S2214-109X(20)30311-9

44. Dieleman JL, Cao J, Chapin A, et al. US Health Care Spending by Payer and Health Condition, 1996-2016. *JAMA*. 2020;323(9):863-884. doi:10.1001/jama.2020.0734

45. Brand AC, Lévy-Piedbois C, Piedbois P, et al. Direct treatment costs for patients with lung cancer from first recurrence to death in france. *Pharmacoeconomics*. 2003;21(9):671-679. doi:10.2165/00019053-200321090-00005

46. Vergnenègre A, Molinier L, Combescure C, Daurès JP, Housset B, Chouaïd C. The Cost of Lung Cancer Management in France from the Payor's Perspective. *Disease Management & Health Outcomes*. 2006;14(1):55-67. doi:10.2165/00115677-200614010-00007

47. Arca JA, Blanco Ramos MÁ, de la Infanta RG, López CP, Pérez LG, López JL. Lung Cancer Diagnosis: Hospitalization Costs. *Archivos de Bronconeumología ((English Edition))*. 2006;42(11):569-574. doi:10.1016/S1579-2129(06)60589-2

48. Fleming I, Monaghan P, Gavin A, O'Neill C. Factors influencing hospital costs of lung cancer patients in Northern Ireland. *Eur J Health Econ*. 2008;9(1):79-86. doi:10.1007/s10198-007-0047-4

49. Cicin I, Oksuz E, Karadurmus N, et al. Economic burden of lung cancer in Turkey: a cost of illness study from payer perspective. *Health Econ Rev*. 2021;11(1):22. doi:10.1186/s13561-021-00322-2

50. Zhang J, Li J, Xiong S, et al. Global burden of lung cancer: implications from current evidence. *Ann Canc Epidemiol*. 2021;5:4-4. doi:10.21037/ace-20-31

51. World Health Organization. Lung cancer. World Health Organization. Published June 26, 2023. Accessed August 4, 2023. https://www.who.int/news-room/fact-sheets/detail/lung-cancer

**Supplementary Material**

**INDEX**

# Epidemiological supporting information

**Table S1. One-year survival probabilities by cancer type and stage used by the model.**

| Type of lung cancer | Stage | One year survival probability literature values* | One year survival probability values adjusted and used by the model | Model assumptions |
|---|---|---|---|---|
| Non-small cell lung cancer | Stage I | 90.00% | 90.00% | Changes in the survival probability of stage III and IV were assumed, as there has been an increase in survival in recent years. In addition, we sought to maintain the ratio with the Globocan data. The assumptions were agreed with the oncological expert. |
| | Stage II | 80.00% | 80.00% | |
| | Stage III | 50.00% | 61.00% | |
| | Stage IV | 25.00% | 40.00% | |
| Small cell lung cancer | Limited stage (I, II, IV) | 63.50% | 63.50% | Changes in the survival probability of the extended stage were assumed, as there has been an increase in survival in recent years. In addition, we sought to maintain the ratio with the Globocan data. The assumptions were agreed with the oncological expert. |
| | Extended stage (IV) | 22.00% | 40.00% | |

**Table S2. Two, three, four and five-year survival probabilities by cancer type and stage used by the model.**

| Type of lung cancer | Stage | Two-year survival probability literature values* | Three-year survival probability literature values* | Four-year survival probability literature values* | Five-year survival probability literature values* |
|---|---|---|---|---|---|
| Non-small cell lung cancer | Stage I | 57,5% | 44,3% | 32,5% | 25,8% |
| | Stage II | 38,5% | 30% | 20% | 16% |
| | Stage III | 25,33% | 17,7% | 13,7% | 10,3% |
| | Stage IV | 11% | 9% | 4% | 3% |
| Small cell lung cancer | Limited stage (I, II, IV) | 37,4% | 27,9% | 21,9% | 17,9% |
| | Extended stage (IV) | 6% | 4% | 3% | 2% |

**Table S3. Years life lost due to premature deaths: total deaths by age group and gender and life expectancy in Argentina**

| Age group (years) | Men | | | Women | | |
|---|---|---|---|---|---|---|
| | Total deaths (n) | Life expectancy (years) | Years life lost due to premature deaths | Total deaths (n) | Life expectancy (years) | Years life lost due to premature deaths |
| < 1 | 0 | 73.51 | 0 | 0 | 79.50 | 0 |
| 1-4 | 0 | 73.19 | 0 | 1 | 79.09 | 79 |
| 5-9 | 1 | 69.27 | 69 | 0 | 75.16 | 0 |
| 10–14 | 1 | 64.33 | 64 | 1 | 70.21 | 70 |
| 15-19 | 3 | 5.40 | 178 | 2 | 65.27 | 131 |
| 20-24 | 2 | 54.66 | 109 | 3 | 60.40 | 181 |
| 25-29 | 3 | 50.02 | 150 | 4 | 55.56 | 222 |
| 30-34 | 17 | 45.34 | 771 | 16 | 50.72 | 812 |
| 35-39 | 27 | 40.65 | 1,098 | 33 | 45.90 | 1,515 |
| 40-44 | 53 | 36.01 | 1,909 | 7 | 41.14 | 288 |
| 45-49 | 103 | 31.43 | 3,237 | 79 | 36.44 | 2,879 |
| 50-54 | 326 | 26.99 | 8,799 | 184 | 31.85 | 5,860 |
| 55-59 | 670 | 22.77 | 15,256 | 370 | 27.37 | 10,127 |
| 60-64 | 1041 | 18.84 | 19,612 | 560 | 23.08 | 12,925 |
| 65-69 | 1291 | 15.27 | 19,714 | 672 | 19.04 | 12,795 |
| 70-74 | 1285 | 12.80 | 16,448 | 636 | 15.22 | 9,680 |
| 75-79 | 985 | 9.30 | 9,161 | 524 | 11.71 | 6,136 |
| 80-84 | 627 | 6.78 | 4,251 | 382 | 8.45 | 3,228 |
| >85 | 446 | 4.67 | 2,083 | 374 | 5.65 | 2,113 |

**Table S4. YLDs disaggregated per gender, cancer type and stage for Argentina, 2023.**

| Type of lung cancer | Stage | N (A1) | N (A2) | Disability Weight (annual) (B) | YLDs - Men (C1) C1 = A1 * B | YLDs - Women (C2) C2 = A2 * B | YLDs - Total |
|---|---|---|---|---|---|---|---|
| Incident cases | | | | | | | |
| Non-small cell | Stage I | 822 | 464 | 0.288 | 237 | 134 | 371 |
| | Stage II | 493 | 279 | 0.288 | 142 | 80 | 222 |
| | Stage III | 1644 | 929 | 0.288 | 474 | 268 | 741 |
| | Stage IV | 3618 | 2044 | 0.451 | 1632 | 922 | 2553 |
| Small cell | Limited stage (I, II, IV) | 406 | 230 | 0.288 | 117 | 66 | 183 |
| | Extended stage (IV) | 754 | 426 | 0.451 | 340 | 192 | 533 |
| Prevalent cases | | | | | | | |
| Non-small cell | Stage I | 2050 | 1372 | 0.049 | 100 | 67 | 168 |
| | Stage II | 908 | 596 | 0.049 | 44 | 29 | 74 |
| | Stage III | 2090 | 1360 | 0.049 | 102 | 67 | 169 |
| | Stage IV | 2417 | 1521 | 0.451 | 1090 | 686 | 1776 |
| Small cell | Limited stage (I, II, IV) | | | | | | |
| | Extended stage (IV) | 731 | 483 | 0.049 | 36 | 24 | 59 |
| Total burden | | | | | | | |
| All cancer types and stages | | 8536 | 5543 | | 4468 | 2629 | 7097 |

**Table S5.** Disability-adjusted life years (DALYs) disaggregated per gender for Argentina, 2023.

| Age group | Men | Woman | Total |
|---|---|---|---|
| YLDs | 4,468 | 2,629 | 7,097 |
| YLLs | 102,908 | 69,040 | 171,949 |
| DALYs | 107,376 | 71,670 | 179,046 |

# Cost supporting information

**Table S6. Unit costs of healthcare resources used in the diagnosis, staging, treatment and follow-up of patients with small and non-small cell lung cancer in Argentina.** Costs are reported in 2023 US$ dollars.

| Health resource | Public sector | Social Security | Private sector | weighted by health sub-sector |
|---|---|---|---|---|
| Consultation with pulmonologist | $ 6.17 | $ 9.69 | $ 11.98 | $ 8.72 |
| Surgical risk consultation | $ 6.17 | $ 9.69 | $ 11.98 | $ 8.72 |
| Consultation with surgeon | $ 6.17 | $ 9.69 | $ 11.98 | $ 8.72 |
| Consultation with oncologist | $ 6.17 | $ 9.69 | $ 11.98 | $ 8.72 |
| Consultation with radiotherapist | $ 6.17 | $ 9.69 | $ 11.98 | $ 8.72 |
| Haemostasis/ coagulogram | $ 3.90 | $ 6.65 | $ 7.57 | $ 5.76 |
| Platelets | $ 0.65 | $ 1.11 | $ 1.26 | $ 0.96 |
| Blood count | $ 1.95 | $ 3.33 | $ 3.79 | $ 2.88 |
| Hepatogram | $ 3.90 | $ 6.65 | $ 7.57 | $ 5.76 |
| Urea | $ 0.98 | $ 1.66 | $ 1.89 | $ 1.44 |
| Creatinine | $ 1.30 | $ 2.22 | $ 2.52 | $ 1.92 |
| Calcium | $ 2.60 | $ 4.44 | $ 5.05 | $ 3.84 |
| LDH | $ 1.95 | $ 3.33 | $ 3.79 | $ 2.88 |
| ALP | $ 0.98 | $ 1.66 | $ 1.89 | $ 1.44 |
| 24-hour urine | $ 1.95 | $ 3.33 | $ 3.79 | $ 2.88 |
| Lung laboratory | $ 16.01 | $ 27.29 | $ 31.06 | $ 23.61 |
| CT-guided puncture | $ 76.79 | $ 130.90 | $ 148.98 | $ 113.23 |
| Videofibronoscopy | $ 50.55 | $ 86.17 | $ 98.07 | $ 74.54 |
| Mediastinoscopy | $ 162.90 | $ 283.98 | $ 316.03 | $ 243.10 |
| IHC | $ 20.58 | $ 33.87 | $ 39.93 | $ 29.79 |
| X-rays | $ 4.48 | $ 7.20 | $ 8.70 | $ 6.41 |
| Chest CT | $ 32.47 | $ 48.31 | $ 63.00 | $ 44.64 |
| ECG | $ 3.07 | $ 4.86 | $ 5.95 | $ 4.35 |
| Abdomen-Pelvis CT | $ 40.63 | $ 60.41 | $ 78.83 | $ 55.84 |
| Centellogram | $ 14.20 | $ 22.52 | $ 27.55 | $ 20.16 |
| Brain CT with contrast | $ 72.78 | $ 115.03 | $ 141.19 | $ 103.16 |
| Brain MRI with contrast | $ 43.51 | $ 74.18 | $ 84.42 | $ 64.16 |
| PET/CT with FPE | $ 13.46 | $ 20.93 | $ 26.12 | $ 18.92 |
| Spirometry | $ 9.50 | $ 14.89 | $ 18.43 | $ 13.41 |
| Lobectomy/ Pneumonectomy/ Segmentectomy | $ 713.64 | $ 1,235.23 | $ 1,384.46 | $ 1,060.90 |
| IMRT | $ 1,491.16 | $ 2,541.82 | $ 2,892.85 | $ 2,198.74 |
| 3D Radiotherapy | $ 1,084.48 | $ 1,848.60 | $ 2,103.89 | $ 1,599.08 |
| Day hospital | $ 62.19 | $ 106.01 | $ 120.65 | $ 91.70 |
| General ward admission | $ 134.52 | $ 237.16 | $ 260.96 | $ 201.96 |
| Intensive care unit admission | $ 232.50 | $ 399.60 | $ 451.05 | $ 344.34 |

**Abbreviations**. IHC, immunohistochemistry. ECG, electrocardiogram. MRI, magnetic resonance imaging. PET, positron emission tomography. CT, computed tomography. FPE, Fluorodeoxyglucose Positron Emission. IMRT, Intensity modulated radiation therapy. LDH, Lactate dehydrogenase. ALP, Alkaline phosphatase.

**Table S7. Expected amounts per patient per year of each health resource used in the diagnosis, staging, treatment and follow-up of patients with small cell lung cancer.** Expected quantities are reported according to phase of the health event.

| Health phases | Health resource | Incident patient | | Prevalent patient | |
|---|---|---|---|---|---|
| | | Limited stage (I, II, IV) | Extended stage (IV) | Limited stage (I, II, IV) | Extended stage (IV) |
| Diagnosis and staging | Consultation with pulmonologist | 3.00 | 2.00 | - | - |
| | Surgical risk consultation | 0.10 | - | - | - |
| | Consultation with surgeon | 0.10 | - | - | - |
| | Consultation with oncologist | 3.00 | 3.00 | - | - |
| | Haemostasis/ coagulogram, Platelets, Blood count, Hepatogram, Urea, Creatinine, Calcium and LDH. | 1.00 | 1.00 | - | - |
| | ALP | - | 1.00 | - | - |
| | 24-hour urine | - | 1.00 | - | - |
| | Lung laboratory | 0.05 | - | - | - |
| | Abdomen-Pelvis CT | 0.50 | 0.80 | - | - |
| | Centellogram | 0.50 | 0.20 | - | - |
| | Brain CT with contrast | 1.00 | 1.00 | - | - |
| | Brain MRI with contrast | 1.00 | 1.00 | - | - |
| | PET/CT with FPE | 1.60 | 1.95 | - | - |
| | Spirometry | 0.80 | 1.00 | - | - |
| Small cell | Consultation with oncologist | 6.00 | 4.50 | 3.15 | 2.40 |
| | Consultation with radiotherapist | 2.55 | 1.20 | - | - |
| | Blood count, Hepatogram, Urea, Creatinine, Calcium, LDH and ALP. | 6.00 | 4.00 | 2.30 | 2.00 |
| | 24-hour urine | 6.00 | 4.00 | 0.80 | 1.60 |
| | Lobectomy/ Pneumonectomy/ Segmentectomy | 0.10 | - | - | - |
| | IHC | 0.10 | - | - | - |

| | | | | |
|---|---|---|---|---|
| General ward admission | 0.50 | - | - | - |
| Intensive care unit admission | 0.30 | - | - | - |
| Day hospital | 3.60 | 2.53 | 0.50 | 1.00 |
| IMRT | 0.03 | - | - | - |
| 3D Radiotherapy | 0.81 | 0.24 | 0.03 | 0.08 |
| Chest CT | - | - | 2.25 | 0.60 |
| Abdomen-Pelvis CT | - | - | 2.25 | 0.60 |
| Centellogram | - | - | 1.50 | 0.40 |

**Abbreviations**. IHC, immunohistochemistry. ECG, electrocardiogram. MRI, magnetic resonance imaging. PET, positron emission tomography. CT, computed tomography. FPE, Fluorodeoxyglucose Positron Emission. IMRT, Intensity modulated radiation therapy. LDH, Lactate dehydrogenase. ALP, Alkaline phosphatase.

**Table S8. Expected amounts per patient per year of each health resource used in the diagnosis, staging, treatment and follow-up of patients with non-small cell lung cancer.** Expected quantities are reported according to phase of the health event.

| Health phases | Health resource | Incident patient | | | | Prevalent patient | | | |
|---|---|---|---|---|---|---|---|---|---|
| | | Stage I | Stage II | Stage III | Stage IV | Stage I | Stage II | Stage III | Stage IV |
| Diagnosis and staging | Consultation with pulmonologist | 3.00 | 3.00 | 3.00 | 5.00 | - | - | - | - |
| | Surgical risk consultation | 1.00 | 1.00 | 0.13 | - | - | - | - | - |
| | Consultation with a surgeon | 3.00 | 3.00 | 1.39 | - | - | - | - | - |
| | Consultation with oncologist | 3.00 | 3.00 | 2.13 | 3.00 | - | - | - | - |
| | Haemostasis/ coagulogram, Platelets, Blood count, Hepatogram, Urea, Creatinine, Calcium and LDH. | 1.00 | 1.00 | 1.00 | 1.00 | - | - | - | - |
| | 24-hour urine | - | - | - | 1.00 | - | - | - | - |
| | Lung laboratory | 0.20 | 0.20 | 0.05 | - | - | - | - | - |
| | CT-guided puncture | 0.40 | 0.40 | 0.50 | 0.80 | - | - | - | - |
| | Videofibronoscopy | 0.90 | 0.90 | 0.50 | 0.20 | - | - | - | - |
| | Mediastinoscopy | 0.20 | 0.20 | - | - | - | - | - | - |
| | IHC | 1.00 | 1.00 | 1.00 | 1.00 | - | - | - | - |
| | X-rays | 1.00 | 1.00 | 1.00 | 1.00 | - | - | - | - |
| | Chest CT | 1.15 | 1.15 | 1.15 | 1.80 | - | - | - | - |
| | ECG | 1.00 | 1.00 | 0.80 | 1.00 | - | - | - | - |
| | Abdomen-Pelvis CT | 0.15 | 0.15 | 0.15 | 0.80 | - | - | - | - |
| | Centellogram | 0.15 | 0.15 | 0.15 | 0.80 | - | - | - | - |
| | Brain CT with contrast | 0.50 | 0.50 | 0.30 | 0.85 | - | - | - | - |
| | Brain MRI with contrast | 0.50 | 0.50 | 0.20 | 0.15 | - | - | - | - |
| | PET/CT with FPE | 0.85 | 0.85 | 0.85 | 0.20 | - | - | - | - |
| | Spirometry | 1.00 | 1.00 | 0.70 | - | - | - | - | - |
| Treatment and follow-up | Consultation with pulmonologist | 1.40 | 1.40 | 0.47 | - | - | - | - | |
| | Consultation with a surgeon | 2.70 | 2.70 | 0.34 | - | 1.35 | 1.35 | 0.75 | |
| | Consultation with oncologist | 6.00 | 8.21 | 6.29 | 12.60 | 3.70 | 3.70 | 3.90 | 15.00 |
| | Consultation with radiotherapist | - | - | 2.30 | - | - | - | - | - |
| | Blood count, Hepatogram, Urea, Creatinine, Calcium, LDH and ALP. | 3.00 | 5.16 | 4.59 | 11.20 | 3.98 | 3.98 | 4.70 | 15.00 |
| | 24-hour urine | - | - | - | 0.30 | 0.39 | 0.39 | 0.53 | 1.00 |
| | Lobectomy/ Pneumonectomy/ Segmentectomy | 0.90 | 0.90 | 0.17 | - | - | - | 0.00 | |
| | IHC | 0.90 | 0.90 | 0.17 | - | - | - | 0.00 | |
| | General ward admission | 4.50 | 4.50 | 1.19 | - | - | - | 0.00 | |
| | Intensive care unit admission | 1.80 | 1.80 | 0.51 | - | - | - | 0.00 | |
| | IMRT | 0.01 | 0.01 | 0.05 | - | - | - | 0.00 | |
| | 3D Radiotherapy | 0.09 | 0.09 | 0.59 | 0.10 | 0.04 | 0.04 | 0.05 | 0.10 |
| | Chest CT | 3.00 | 2.00 | 1.47 | 2.10 | 2.51 | 2.51 | 2.33 | 3.00 |
| | Abdomen-Pelvis CT | 3.00 | 2.00 | 1.47 | 2.10 | 2.51 | 2.51 | 2.33 | 3.00 |
| | Centellogram | 1.50 | 1.50 | 1.47 | 2.10 | 2.06 | 2.06 | 2.08 | 3.00 |
| | Brain CT with contrast | - | - | - | 0.07 | 0.04 | 0.04 | 0.05 | 0.07 |
| | Brain MRI with contrast | - | - | - | 0.07 | 0.04 | 0.04 | 0.05 | 0.07 |
| | Day hospital | - | - | 4.37 | 4.03 | 1.18 | 1.18 | 1.72 | 4.78 |

**Abbreviations**. IHC, immunohistochemistry. ECG, electrocardiogram. MRI, magnetic resonance imaging. PET, positron emission tomography. CT, computed tomography. FPE, Fluorodeoxyglucose Positron Emission. IMRT, Intensity modulated radiation therapy. LDH, Lactate dehydrogenase. ALP, Alkaline phosphatase.

**Table S9. Costs of treatment per incident patient with non-small cell lung cancer per year.** Dosage, presentation form, wholesale price of drugs, and total drug cost per patient-year. Costs are reported in 2023 US$ dollars.

| Drugs | Dosage | Presentation | Wholesale price per mg* | Cost per patient-year |
|---|---|---|---|---|
| **Stage II (made up of 58% drug costs)** | | | | |
| *Carboplatin + pemetrexed (39%) - First line* | | | | |
| Carboplatin | 600 mg per cycle | 150mg Iny. Amp. X 1 | $ 0.78 | $ 1,567.79 |
| Pemetrexed | 500 mg/m2 per cycle | 500mg F.Amp. X 1 | $ 4.30 | |
| *Cisplatin + pemetrexed (35%) - First line* | | | | |
| Cisplatin | 75 mg/m2 per cycle | 50mg Iny. F.Amp. X 1 | $ 0.64 | $ 1,420.45 |
| Pemetrexed | 500 mg/m2 per cycle | 500mg F.Amp. X 1 | $ 4.30 | |
| *Other monotherapies and drug combinations that account for less than 10% of the stage's drug cost (26%)* | | | | |
| **Stage III (made up of 86% drug costs)** | | | | |
| *Durvalumab (39%) - First line* | | | | |
| Durvalumab | 10 mg/kg every 15 days | 500mg IV Iny X 1 X 10ml | $ 6.53 | $ 5,846.10 |
| *Carboplatin + pemetrexed (15%) - First line* | | | | |
| Carboplatin | 600 mg per cycle | 150mg Iny. Amp. X 1 | $ 0.78 | $ 2,203.88 |
| Pemetrexed | 500 mg/m2 per cycle | 500mg F.Amp. X 1 | $ 4.30 | |
| *Carboplatin + paclitaxel (13%) - First line* | | | | |
| Carboplatin | 600 mg per cycle | 150mg Iny. Amp. X 1 | $ 0.78 | $ 1,904.38 |
| Paclitaxel | 175 mg/m2 per cycle | 150mg Iny. Amp. X 1 | $ 2.24 | |
| *Cisplatin + pemetrexed (12%) - First line* | | | | |
| Cisplatin | 75 mg/m2 per cycle | 50mg Iny. F.Amp. X 1 | $ 0.64 | $ 1,772.89 |
| Pemetrexed | 500 mg/m2 per cycle | 500mg F.Amp. X 1 | $ 4.30 | |
| *Other monotherapies and drug combinations that account for less than 10% of the stage's drug cost (21%)* | | | | |
| **Stage IV (made up of 96% drug costs)** | | | | |
| *Carboplatin + pemetrexed + pembrolizumab (21%) - First line* | | | | |
| Carboplatin | 600 mg per cycle | 150mg Iny. Amp. X 1 | $ 0.78 | $ 7,592.47 |
| Pemetrexed | 500 mg/m2 per cycle | 500mg F.Amp. X 1 | $ 4.30 | |
| Pembrolizumab | 400 mg every 6 weeks | 100mg Vial X 1 X 4ml | $ 32.62 | |
| *Pembrolizumab (16%) - First line* | | | | |
| Pembrolizumab | 400 mg every 6 weeks | 100mg Vial X 1 X 4ml | $ 32.62 | $ 5,774.00 |
| *Other monotherapies and drug combinations that account for less than 10% of the stage's drug cost (63%)* | | | | |

*Wholesale price is estimated as the retail price (RRP) divided by 1.7545.
Source: Kairos, Alfabeta, ANMAT disposition 10071, ANMAT disposition 7642, ANMAT disposition 9057, ANMAT disposition 5800, ANMAT disposition 1785, ANMAT disposition 38887.

**Table S10. Costs of treatment per incident patient with small cell lung cancer per year**. Dosage, presentation form, wholesale price of drugs, and total drug cost per patient-year. Costs are reported in 2023 US$ dollars.

| Drugs | Dosage | Presentation | Wholesale price per mg* | Cost per patient-year |
|---|---|---|---|---|
| **Limited stage (I, II, IV) (made up of 37% drug costs)** | | | | |
| *Carboplatin + etoposide (20%) - First line* | | | | |
| Carboplatin | 600 mg per cycle | 150mg Iny. Amp. X 1 | $ 0.78 | $ 914.90 |
| Etoposide | 300 mg/m2 per cycle | 100mg IV Iny X 1 X 5ml | $ 0.08 | |
| *Cisplatin + etoposide (80%) - First line* | | | | |
| Cisplatin | 75 mg/m2 per cycle | 50mg Iny. F.Amp. X 1 | $ 0.64 | $ 221.69 |
| Etoposide | 300 mg/m2 per cycle | 100mg IV Iny X 1 X 5ml | $ 0.08 | |
| **Extended stage (made up of 72% drug costs)** | | | | |
| *Carboplatin + etoposide + Atezolizumab (73%) - First line* | | | | |
| Carboplatin | 600 mg per cycle | 150mg Iny. Amp. X 1 | $ 0.78 | $ 1,988.48 |
| Etoposide | 300 mg/m2 per cycle | 100mg IV Iny X 1 X 5ml | $ 0.08 | |
| Atezolizumab | 1200 mg per cycle | 1200mg Vial X 1 X 20ml | $ 4.08 | |
| *Carboplatin + etoposide (24%) - First line* | | | | |
| Carboplatin | 600 mg per cycle | 150mg Iny. Amp. X 1 | $ 0.78 | $ 648.05 |
| Etoposide | 300 mg/m2 per cycle | 100mg IV Iny X 1 X 5ml | $ 0.08 | |

*Other monotherapies and drug combinations that account for less than 10% of the stage's drug cost (3%)*
*Wholesale price is estimated as the retail price (RRP) divided by 1.7545.
Source: Kairos, Alfabeta, ANMAT disposition 10071, ANMAT disposition 3034, ANMAT disposition 3055, ANMAT disposition 38887.

**Table S11. Costs of treatment per prevalent patient with non-small cell lung cancer per year.** Dosage, presentation form, wholesale price of drugs, and total drug cost per patient-year. Costs are reported in 2023 US$ dollars.

| Drugs | Dosage | Presentation | Wholesale price per mg* | Cost per patient-year |
|---|---|---|---|---|
| **Stage I and II (made up of 83% drug costs)** | | | | |
| *Osimertinib (45%) - First line* | | | | |
| Osimertinib | 80 mg per day | 800mg comp. X 30 | $ 2.20 | $ 868.34 |
| *Carboplatin + pemetrexed + pembrolizumab (11%) - First line* | | | | |
| Carboplatin | 600 mg per cycle | 150mg Iny. Amp. X 1 | $ 0.78 | $ 3,792.43 |
| Pemetrexed | 500 mg/m2 per cycle | 500mg F.Amp. X 1 | $ 4.30 | |
| Pembrolizumab | 400 mg every 6 weeks | 100mg Vial X 1 X 4ml | $ 32.62 | |
| *Other monotherapies and drug combinations that account for less than 10% of the stage's drug cost (44%)* | | | | |
| **Stage III (made up of 93% drug costs)** | | | | |
| *Durvalumab (25%) - First line* | | | | |
| Durvalumab | 10 mg/kg every 15 days | 500mg IV Iny X 1 X 10ml | $ 6.53 | $ 1,949.50 |
| *Carboplatin + pemetrexed + pembrolizumab (16%) - First line* | | | | |
| Carboplatin | 600 mg per cycle | 150mg Iny. Amp. X 1 | $ 0.78 | $ 3,792.43 |
| Pemetrexed | 500 mg/m2 per cycle | 500mg F.Amp. X 1 | $ 4.30 | |
| Pembrolizumab | 400 mg every 6 weeks | 100mg Vial X 1 X 4ml | $ 32.62 | |
| *Pembrolizumab (12%) - First line* | | | | |
| Pembrolizumab | 400 mg every 6 weeks | 100mg Vial X 1 X 4ml | $ 32.62 | $ 2,883.74 |
| *Other monotherapies and drug combinations that account for less than 10% of the stage's drug cost (47%)* | | | | |
| **Stage IV (made up of 96% drug costs)** | | | | |
| *Pembrolizumab (28%) - Second line* | | | | |
| Pembrolizumab | 400 mg every 6 weeks | 100mg Vial X 1 X 4ml | $ 32.62 | $ 7,870.91 |
| *Carboplatin + pemetrexed + pembrolizumab (13%) - First line* | | | | |
| Carboplatin | 600 mg per cycle | 150mg Iny. Amp. X 1 | $ 0.78 | $ 3,699.38 |
| Pemetrexed | 500 mg/m2 per cycle | 500mg F.Amp. X 1 | $ 4.30 | |
| Pembrolizumab | 400 mg every 6 weeks | 100mg Vial X 1 X 4ml | $ 32.62 | |
| *Pembrolizumab (10%) - First line* | | | | |
| Pembrolizumab | 400 mg every 6 weeks | 100mg Vial X 1 X 4ml | $ 32.62 | $ 2,883.74 |
| *Other monotherapies and drug combinations that account for less than 10% of the stage's drug cost (49%)* | | | | |

*Wholesale price is estimated as the retail price (RRP) divided by 1.7545.
Source: Kairos, Alfabeta, ANMAT disposition 7144, ANMAT disposition 10071, ANMAT disposition 7642, ANMAT disposition 1785, ANMAT disposition 9057.

**Table S12. Costs of treatment per prevalent patient with small cell lung cancer per year.** Dosage, presentation form, wholesale price of drugs, and total drug cost per patient-year. Costs are reported in 2023 US$ dollars.

| Drugs | Dosage | Presentation | Wholesale price per mg* | Cost per patient-year |
|---|---|---|---|---|
| **Limited stage (I, II, IV) (made up of 19% drug costs)** | | | | |
| *Carboplatin + atezolizumab + etoposide (11%) - First line* | | | | |
| Carboplatin | 600 mg per cycle | 150mg Iny. Amp. X 1 | $ 0.78 | $ 424.21 |
| Atezolizumab | 1200 mg per cycle | 1200mg Vial X 1 X 20ml | $ 4.08 | |
| Etoposide | 300 mg/m2 per cycle | 100mg IV Iny X 1 X 5ml | $ 0.08 | |
| *Other monotherapies and drug combinations that account for less than 10% of the stage's drug cost (89%)* | | | | |
| **Extended stage (made up of 52% drug costs)** | | | | |
| *Paclitaxel (46%) - First line* | | | | |
| Paclitaxel | 175 mg/m2 per cycle | 150mg Iny. Amp. X 1 | $ 2.24 | $ 249.84 |
| *Carboplatin + etoposide (23%) - First line* | | | | |
| Carboplatin | 600 mg per cycle | 150mg Iny. Amp. X 1 | $ 0.78 | $ 127.07 |
| Etoposide | 300 mg/m2 per cycle | 100mg IV Iny X 1 X 5ml | $ 0.08 | |
| *Other monotherapies and drug combinations that account for less than 10% of the stage's drug cost (31%)* | | | | |

*Wholesale price is estimated as the retail price (RRP) divided by 1.7545.
Source: Kairos, Alfabeta, ANMAT disposition 10071, ANMAT disposition 3055, ANMAT disposition 3034, ANMAT disposition 5800.

**Table S13. Adverse event management costs included in annual cost per incident and prevalent lung cancer patient.** Costs are reported in 2023 US$ dollars.

| Regimen | Public sector | Social security | Private sector | weighted by health sub-sector |
|---|---|---|---|---|
| Anemia | $ 1009.96 | $ 1587.95 | $ 1739.04 | $ 1392.49 |
| Neutropenia | $ 23.51 | $ 37.69 | $45.60 | $ 33.57 |
| Plateletopenia | $ 196.18 | $ 341.01 | $ 380.59 | $ 292.31 |
| Pneumonia | $ 2754.46 | $ 3836.39 | $ 4151.32 | $ 3475.64 |
| Diarrhoea | $ 841.53 | $ 1098.33 | $ 1186.30 | $ 1014.82 |
| Vomiting | $ 529.77 | $ 827.72 | $ 908.92 | $ 727.49 |
| Nausea | $ 529.77 | $ 827.72 | $ 908.92 | $ 727.49 |
| Pneumonitis | $ 4638.11 | $ 5502.74 | $ 5749.08 | $ 5213.59 |

**Table S14. Adverse event prevalence rate.**

| Regimen | ALK | Anti EGFR | Immunotherapy | Chemotherapy |
|---|---|---|---|---|
| Anemia | 3.00% | 1.35% | 1.98% | 6.85% |
| Neutropenia | 2.65% | 0.10% | 0.00% | 15.60% |
| Plateletopenia | 0.00% | 0.25% | 0.00% | 1.85% |
| Pneumonia | 3.30% | 2.25% | 1.70% | 3.43% |
| Diarrhoea | 1.35% | 5.85% | 1.55% | 0.68% |
| Vomiting | 1.65% | 1.28% | 0.15% | 0.55% |
| Nausea | 2.00% | 0.23% | 0.20% | 1.40% |
| Pneumonitis | 1.00% | 0.25% | 3.10% | 0.18% |

Serious adverse events were only included if they scored above 3 on the Common Terminology Criteria for Adverse Effects V 5.0 (CTCAE) scale and if they occurred in more than 3% of patients on some of the treatment regimens.

**Table S15. Unit costs of healthcare resources used in the estimated cost of death for prevalent and incident small cell and non-small cell lung cancer patients.** Costs are reported in 2023 US$ dollars.

| Health resource | Public sector | Social Security | Private sector | weighted by health sub-sector |
|---|---|---|---|---|
| Palliative care | $ 62.19 | $ 106.01 | $ 120.65 | $ 91.70 |
| General ward admission | $ 134.52 | $ 237.16 | $ 260.96 | $ 201.96 |

**Table S16. Expected amounts of each health resource used in the estimated cost of death for prevalent and incident small cell and non-small cell lung cancer patients.**

| Health resource | Expected amounts |
|---|---|
| Palliative care | 0.40 |
| General ward admission | 4.80 |

These amounts are in addition to the resources coming from the stages imputed in the cost of mortality.

**Table S17. Average annual cost per incident lung cancer patient in Argentina.** Costs are reported in 2023 US$ dollars.

| Type of lung cancer | Stage | Public sector | Social Security | Private sector | weighted by health sub-sector |
|---|---|---|---|---|---|
| Non-small cell | Stage I | $2,478 | $4,196 | $4,807 | $3,641 |
| | Stage II | $7,657 | $9,560 | 10,213 | $8,941 |
| | Stage III | $17,220 | $18,574 | $19,081 | $18,140 |
| | Stage IV | $36,790 | $37,603 | $37,987 | $37,356 |
| | Subtotal | $25,424 | $26,567 | $27,030 | $26,206 |
| Small cell | Limited stage (I, II, IV) | $3,117 | $4,437 | $4,918 | $4,012 |
| | Extended stage (IV) | $3,710 | $4,318 | $4,578 | $4,129 |
| | Subtotal | $3,503 | $4,359 | $4,697 | $4,088 |
| Total | | $22,135 | $23,236 | $23,680 | $22,889 |

**Table S18. Average annual cost per prevalent lung cancer patient in Argentina.** Costs are reported in 2023 US$ dollars.

| Type of lung cancer | Stage | Public sector | Social Security | Private sector | weighted by health sub-sector |
|---|---|---|---|---|---|
| Non-small cell | Stage I | $10,906 | $11,182 | $11,337 | $11,102 |
| | Stage II | $10,906 | $11,182 | $11,337 | $11,102 |
| | Stage III | $14,721 | $15,042 | $15,209 | $14,947 |
| | Stage IV | $29,688 | $30,421 | $30,752 | $30,195 |
| | Subtotal | $17,971 | $18,405 | $18,620 | $18,275 |
| Small cell | Limited stage (I, II, IV) | $549 | $725 | $838 | $676 |
| | Extended stage (IV) | $633 | $810 | $885 | $755 |
| | Subtotal | $575 | $751 | $853 | $701 |
| Total | | $15,710 | $16,110 | $16,310 | $15,990 |

**Table S19. Cost of death per incident patient by health subsector.** Costs are reported in 2023 US$ dollars.

| Type of lung cancer | Stage | Public sector | Social Security | Private sector | weighted by health sub-sector |
|---|---|---|---|---|---|
| Non-small cell and small cell | All stages | $645.68 | $1,138.38 | $1,252.61 | $969.43 |

**Table S20. Cost of death per prevalent patient by health subsector for each type and stage of lung cancer in Argentina.** Costs are reported in 2023 US$ dollars.

| Type of lung cancer | Stage | Public sector | Social Security | Private sector | weighted by health sub-sector |
|---|---|---|---|---|---|
| Non-small cell | Stage I | $5,267.75 | $6,021.30 | $6,290.93 | $5,778.10 |
| | Stage II | $5,267.75 | $6,021.30 | $6,290.93 | $5,778.10 |
| | Stage III | $6,793.77 | $7,565.61 | $7,839.56 | $7,316.14 |
| | Stage IV | $12,780.17 | $13,717.10 | $14,056.41 | $13,415.35 |
| Small cell | Limited stage (I, II, IV) | $1,125.00 | $1,838.56 | $2,091.42 | $1,607.87 |
| | Extended stage (IV) | $1,158.36 | $1,872.73 | $2,110.34 | $1,639.29 |

**Table S21. Economic burden of lung cancer by subsector of the health system in Argentina. Cost are express in US$**

| | Public sector | Social Security | Private sector | Total |
|---|---|---|---|---|
| **Incidents cases** | | | | |
| **Non-small cell lung cancer** | | | | |
| Stage I | $1.211.544 | $2.483.501 | $989.640 | $4.684.685 |
| Stage II | $2.246.440 | $3.394.958 | $1.261.525 | $6.902.923 |
| Stage III | $16.838.822 | $21.986.884 | $7.856.459 | $46.682.165 |
| Stage IV | $79.148.474 | $97.927.774 | $34.409.958 | $211.486.206 |
| **Small cell lung cancer** | | | | |
| Limited stage (I, II, IV) | $753.044,04 | $1.297.537 | $500.308 | $2.550.889 |
| Extended stage (IV) | $1.664.745 | $2.345.102 | $864.779 | $4.874.626 |
| **Subtotal cost for incident cases** | $101.863.069 | $129.435.755 | $45.882.670 | $277.181.494 |
| **Subtotal cost of incident death** | $ 1.390.423 | $ 2.967.518 | $ 1.135.756 | $ 5.493.698 |
| **Total** | **$103.253.493** | **$132.403.273** | **$47.018.426** | **$282.675.192** |
| **Prevalents cases** | | | | |
| **Non-small cell lung cancer** | | | | |
| Stage I | $14.177.328 | $17.595.335 | $6.205.108 | $37.977.771 |
| Stage II | $6.231.681 | $7.734.075 | $2.727.471 | $16.693.226 |
| Stage III | $19.448.353 | $24.056.068 | $8.459.818 | $51.964.239 |
| Stage IV | $44.414.286 | $55.092.268 | $19.370.965 | $118.877.519 |
| **Small cell lung cancer** | | | | |
| Limited stage (I, II, IV) | $253.327 | $404.642 | $162.708 | $820.676 |
| Extended stage (IV) | $132.673 | $205.624 | $78.165 | $416.462 |
| **Subtotal** | **$84.657.647** | **$105.088.011** | **$37.004.235** | **$226.749.893** |
| **Death after one year of diagnosis** | **$ 15.873.696** | **$ 22.606.352** | **$ 8.283.431** | **$ 46.763.479** |
| **Total costo of lung cancer in Argentina** | **$203.784.836** | **$260.097.636** | **$92.306.092** | **$556.188.564** |